\documentstyle[prd,aps,floats,tabularx,epsfig]{revtex}
\newcommand{\be}{\begin{equation}}
\newcommand{\ee}{\end{equation}}
\newcommand{\bea}{\begin{eqnarray}}
\newcommand{\eea}{\end{eqnarray}}

\newcommand{\pp}{~~~.}
\newcommand{\vv}{~~~,}

\newcommand{\Nature}{{\it Nature\,}}

\newcommand{\ApJ}{{\it Astrophys. J.\,}}

\newcommand{\NP}{{\it Nucl. Phys.\,}}

\begin{document}
\setlength{\unitlength}{1mm}
\twocolumn[\hsize\textwidth\columnwidth\hsize\csname@twocolumnfalse\endcsname

\title{{\small{\hfill $\begin{array}{r} \mbox{DSF 25/2000, astro-ph/0007419}
\end{array}$}}\\
Testing Standard and Degenerate Big Bang Nucleosynthesis with BOOMERanG and
MAXIMA-1}
\author{S. Esposito$^1$, G. Mangano$^{1}$, A. Melchiorri$^{2}$, G.
Miele$^{1}$, O. Pisanti$^{1}$}
\address{$^1$
Dipartimento di Scienze Fisiche, Universit\'{a} ``Federico II'', Napoli,
and INFN Sezione di Napoli, Italy.\\
$^2$ Dipartimento di Fisica, Universit\'{a} ``La Sapienza'', Roma, Italy.\\}

\maketitle

\begin{abstract}

We test different Big Bang Nucleosynthesis scenarios using the recent
results on the Cosmic Microwave Background Anisotropies provided by the
BOOMERanG and MAXIMA-1 experiments versus the observed abundances of
$^4He$, $D$ and $^7Li$. The likelihood analysis, based on Bayesian
approach, shows that, in the case of high deuterium abundance,
$Y_D=(2.0{\pm}0.5)10^{-4}$, both standard and degenerate BBN are
inconsistent with the CMBR measurement at more than $3
\sigma$. Assuming low deuterium abundance, $Y_D=(3.4{\pm}0.3)10^{-5}$, the
standard BBN model is still inconsistent with present observations at $
2 \sigma$ level, while the degenerate BBN results to be compatible.
 Unless systematics effects will be found in nuclide abundances
and/or in CMBR data analysis this result may be a signal in favour of new
physics like a large chemical potential of the relic neutrino-antineutrino
background.
\end{abstract}
\bigskip
%
]
One of the main goals of modern cosmology is the knowledge of the energy
density content of the universe. The four parameters $\Omega_B$,
$\Omega_{CDM}$, $\Omega_{\nu}$ and $\Omega_\Lambda$, giving, respectively,
the baryon, cold dark matter, neutrino and cosmological constant
contributions to the total energy density, in unit of the critical density,
and the Hubble constant, $H_0=100h ~Km ~s^{-1} Mpc^{-1}$, enter as crucial
parameters in several cosmological observables. A well known example is
provided by the baryon density parameter which plays an essential role in
determining the abundances of light nuclides produced in the early
universe. An increasing interest has been also devoted to other issues, as
structure formation and the anisotropy of the Cosmic Microwave Background
Radiation (CMBR). As for our theoretical understanding of the Big Bang
Nucleosynthesis (BBN), to test the theoretical models which describe these
aspects of the hot Big Bang model, it is essential to have precise
measurements of the several $\Omega_ih^2$. Furthermore, combining different
cosmological observables, and comparing the way they are able to constrain
the $\Omega_ih^2$, allows for a check of the consistency of our present
understanding of the evolution of the universe. This can provide new hints
on phenomena which took place at the macroscopic cosmological level, or
rather related with the very microscopic structure of fundamental
interactions.

This letter represents a contribution in this direction. In particular, we
do perform a combined analysis of the dependence on the energy fractions
$\Omega_Bh^2$ and $\Omega_\nu h^2 = N_{\nu} \Omega_\nu^0 h^2$ for {\it
massless} neutrinos ($N_\nu$ standing for the effective neutrino number and
$\Omega_\nu^0 h^2$ for the energy contribution of a single
$\nu-\overline{\nu}$ specie) of CMBR anisotropies and BBN. This is aimed to
test the standard and degenerate BBN scenario, using the recent results of
the BOOMERanG \cite{Boomerang} and MAXIMA-1 \cite{Maxima} CMBR experiments
and the measurements of $^4He$, $D$ and $^7Li$ primordial abundances.

The theoretical tools necessary to achieve this goal are nowadays rather
robust and the accuracy in the predictions of the Big Bang model is
remarkably improved. In fact, recently, a new generation of BBN codes have
been developed \cite{Emmp1,Emmp2,altri1,altri2}, which give the $^4He$ mass
fraction $Y_p$ with a theoretical error of few per mille.  This effort is
justified in view of the small statistical error which is now quoted in the
$Y_p$ measurements. On the other hand, the theoretical predictions on the
CMBR anisotropies angular power spectrum, in the case of primordial {\it
passive} and {\it coherent} perturbations, expected in the {\it most
general} inflationary model, have also recently reached a $1 \%$ level
accuracy, with a refined treatment of H, He I, and He II
recombination\cite{seag}.

Concerning the CMBR experimental data, after the COBE satellite first
detection of CMBR anisotropies, at scales larger than $5^o$, and more than
$20$ independent detections at different frequencies and scales, see e.g.
\cite{pier}, an important new insight is represented by the recent results
obtained by the BOOMERanG Collaboration \cite{Boomerang}. For the first
time, in fact, multifrequency maps of the microwave background anisotropies
were realized over a significant part of the sky, with $\sim 10'$
resolution and high signal to noise ratio. The anisotropy power spectrum,
$C_\ell$, was measured in a wide range of angular scales from multipole
$\ell \sim 50$ up to $\ell \sim 600$, with error bars of the order of $10
\%$, showing a peak at $\ell_{peak}=(197 {\pm} 6)$ with an amplitude
$DT_{200}=(69 {\pm} 8)\mu K$. While the presence of such peak, compatible
with inflationary scenario, was already suggested by previous measurements
\cite{b97}, the absence of secondary peaks after $\ell \ge 300$ with a flat
spectrum with an amplitude of $\sim 40 \mu K$ up to $\ell \sim 625$ was a
new and unexpected result. This result obtained then an impressive
confirmation by the MAXIMA-1 \cite{Maxima} experiment up to $\ell \sim 800$.

As far as BBN is concerned, in the last few years many results have been
also obtained on light element primordial abundances. The $^4He$ mass
fraction $Y_p$, has been measured with a $0.1 \%$ precision in two
independent surveys, from regression to zero metallicity in Blue Compact
Galaxies, giving a {\it low} value $Y_p^{(l)} = 0.234 {\pm} 0.003$
\cite{lowhe}, and a high one $Y_p^{(h)} = 0.244 {\pm} 0.002$ \cite{highhe},
which are compatible at $2\sigma$ level only, may be due to large
systematic errors. As in \cite{Emmp2}, in our analysis we adopt the more
conservative value $Y_p = 0.238 {\pm} 0.005$.\\ A similar controversy holds
in $D$ measurements, where observations in different Quasars Absorption
line Systems (QAS) lead to the incompatible results $Y_D^{(l)} = \left(3.4
{\pm} 0.3 \right) 10^{-5}$ \cite{lowd}, and $Y_D^{(h)} = \left(2.0 {\pm}
0.5 \right) 10^{-4}$ \cite{highd}. We will perform our analysis for both
low and high $D$ data. Finally, the most recent estimate for $^7Li$
primordial abundance, from the {\it Spite plateau} observed in the halo of
POP II stars, gives $Y_{^7Li} =\left(1.73 {\pm} 0.21 \right) 10^{-10}$
\cite{Li7}. The light nuclide yields strongly depend on the baryon matter
content of the universe, $\Omega_B h^2$. High values for this parameter
result in a larger $^4He$ mass fraction and a lower deuterium number
density. In particular, assuming a standard BBN scenario, i.e. vanishing
neutrino chemical potentials, the likelihood analysis gives, at $95 \%$
C.L. \cite{Emmp2},
\be
\begin{array}{lcc} \mbox{low D} &
\Omega_B h^2 = .017 {\pm} .003 & 1.7 \leq N_\nu \leq 3.3 \vv \\
\mbox{high D} &
\Omega_B h^2 = .007^{+.007}_{-.002}&  2.3 \leq N_\nu \leq 4.4 \pp
\end{array}
\label{standard}
\ee
As already pointed out by several authors \cite{lange,crisis?,crisis?deg},
these values for $\Omega_B h^2$, though in the correct order of magnitude,
are however somehow smaller than the baryon fraction which more easily fit
the CMBR data. In fact, as mentioned, while the narrow first peak around $l
\sim 200$ is a confirmation of the inflationary paradigma, a flat universe
with adiabatic perturbations, the lack of observation of a secondary peak at
smaller scales raises new intriguing questions about the values of the
cosmological parameters. In particular, this result may be a signal in
favour of a larger $\Omega_B h^2 \sim 0.03$, since increasing the baryon
fraction enhances the odd peaks only. In this respect the measurement of
the third peak at larger multipole moments is extremely important. Though
it is fair to say that a further analysis of the BOOMERanG and MAXIMA-1
data, as well as new data from future experiment, are needed to clarify this
issue, it is however timely to study how our theoretical understanding of
BBN can be reconciled with the larger values of $\Omega_B h^2$ suggested by
CMBR data.

In figure 1 we start addressing the problem quantitatively, by reporting our
theoretical predictions on the several abundances, together with the
corresponding experimental values, and the CMBR 68$\%$ C.L. bound on
$\Omega_B h^2=0.030^{+0.007}_{-0.004}$, obtained by our joint analysis of
both BOOMERanG and MAXIMA-1 data (see below for the details). The BBN
predictions are obtained in the standard scenario for $N_\nu=3$.
\begin{figure}[ht]
\begin{center}
\epsfxsize=6cm
\epsfysize=9.8cm
\epsffile{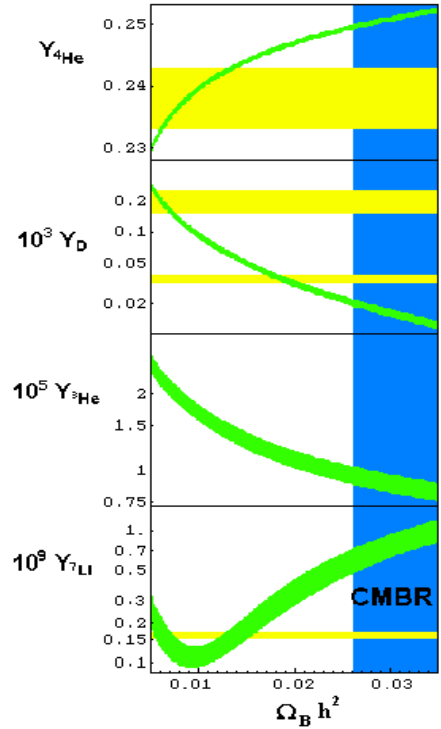}
\end{center}
\caption{The theoretical abundances for $D$, $^3He$, $^4He$ and $^7Li$
nuclides, for $N_\nu=3$, versus the experimental results, represented by the
horizontal bands, are reported. The long vertical band represents the
68$\%$ C.L. interval for $\Omega_B h^2$ obtained by the BOOMERanG and
MAXIMA-1 data analysis.}
\end{figure}
Unless systematics effects will be found in nuclide abundances and/or in
CMBR data analysis, we see that there is a rather sensible disagreement
between the values of the baryon energy density required by standard BBN
and by CMBR data. In particular, from the 68$\%$ C.L. CMBR result for
$\Omega_B h^2$, we can derive the following constraints on the $Y_i$,
\bea
0.2487 < &Y_{^4He}& < 0.2536 \nonumber\\
9.23 {\cdot} 10^{-6} < &Y_D& < 2.32 {\cdot} 10^{-5}\nonumber\\
7.92 {\cdot} 10^{-6}<&Y_{^3He}&<9.85 {\cdot} 10^{-6}\\
5.45 {\cdot} 10^{-10} < &Y_{^7Li}& < 1.25 {\cdot} 10^{-9} \vv
\nonumber
\eea
which are systematically different from the available measured values. We
stress once again that our theoretical analysis is accurate, at worst, at
the level of few percent\footnote{It is important to note that this result
is obtained following a Bayesian approach in the analysis. Assuming the
standard BBN, the best fit model for the CMBR spectrum has a normalized
$\chi^2 \sim 50$, over $\sim 40$ degrees of freedom. Being the $\chi^2$
slightly greater than one, a CMB analysis using the BBN prior as in
\cite{lange} is still perfectly consistent.}.

It has already been stressed \cite{crisis?deg,Emmp2} that a simple way to
improve the agreement of observed nuclide abundances with $\Omega_B h^2
\geq 0.02$ is to assume non vanishing neutrino chemical potentials at the
BBN epoch, a scenario already extensively studied in the past
\cite{KangSteigman}. The effect of neutrino chemical potentials
$\mu_\alpha$, with $\alpha$ the neutrino specie, is twofold. A
non-vanishing $\xi_\alpha=\mu_{\nu_\alpha}/T_\nu$, contribute to $N_\nu$ as
\begin{equation} N_{{\nu}} = 3 + \Sigma_{\alpha} \left[ \frac{30}{7}
\left( \frac{\xi_\alpha}{\pi} \right)^2 +
\frac{15}{7} \left( \frac{\xi_\alpha}{\pi} \right)^4 \right] \vv \label{neff1}
\end{equation}
implying a larger expansion rate of the universe with respect to the
non-degenerate scenario, and a higher value for the neutron to proton
density ratio at the freeze-out. Furthermore, a positive (negative) value
for $\xi_e$ means a larger (smaller) number of $\nu_e$ with respect to
$\bar{\nu}_e$, thus enhancing (lowering) $n \rightarrow p$ processes.
Notice that extra relativistic degrees of freedom, like light sterile
neutrinos, would contribute to $N_\nu$ as well, and in this respect BBN
cannot distinguish between their contribution to the total universe
expansion rate and the one due to neutrino degeneracy. Therefore our
estimates for $N_\nu$ can only represent an upper bound for the total
neutrino chemical potentials. However, it is also worth stressing that a
large $N_\nu >3 $, $does$ indeed require a positive $\xi_e$, since to end
up with the observed values for the $Y_i$, a fine tuned balance between
$\xi_e$ and $N_\nu$ is required \cite{KangSteigman,Emmp2}.

Increasing $N_\nu$ also weakly affects the CMBR anisotropy spectrum  in two
ways. The growth of perturbations inside the horizon is in fact lowered,
resulting in a decay of the gravitational potential and hence in an
increase of the anisotropy near the first peak. Moreover, the size of
horizon and sound horizon at the last scattering surface is changed, and
this, with additional effects in the damping, varies the amplitude and
position of the other peaks, see e.g. \cite{hu}.

To test the degenerate BBN scenario we have performed a likelihood analysis
of the data. First, to constrain the values of the parameter set $(\xi_e$,
$N_\nu$, $\Omega_B h^2)$ from the data on $^4He$, $D$ and $^7Li$ we define
a {\it total likelihood function},
\be
{\cal L}_{Nucl}(N_\nu,\Omega_B h^2,\xi_e)
= L_D ~L_{^4He} ~ L_{^7Li}~~~,
\ee
as described in Ref.\cite{Emmp2}. For a fully degenerate BBN, since the
effect of a positive $\xi_e$ can be compensated by larger $N_\nu$, ${\cal
L}_{Nucl}(N_\nu,\Omega_B h^2,\xi_e)$ may sensibly differ from zero in a
region with rather large values of $N_\nu$. We have chosen to constrain
this parameter to be $N_\nu <16$. This upper limit has been considered after
checking that it is well outside the $95 \%$ upper limit on $N_\nu$ from
the BOOMERanG and MAXIMA-1 data (again, see below). The other two
parameters are chosen in the following ranges, $-1\leq\xi_e\leq1$ and
$0.004\leq\Omega_B h^2\leq0.110$.

Since CMBR spectrum is not sensible to $\xi_e$ alone, we have marginalized
over $\xi_e$. All nuclide abundances have been evaluated using the new BBN
code described in \cite{Emmp2}, while the theoretical uncertainties
$\sigma_i^{th\, 2} (N_\nu,\, \Omega_B h^2)$ are found by linear propagation
of the errors affecting the various nuclear rates entering in the
nucleosynthesis reaction network \cite{altri2}.

The CMBR data analysis methods have been already extensively described in
various papers \cite{lange}. The anisotropy power spectrum, was estimated
in $12$ orthogonalized (independent) bins between $\ell= 50$ and $\ell=
650$ by the BOOMERanG experiment and in $10$ bins, from $\ell= 70$ to $\ell
=750$, by the MAXIMA-1 experiment.  The likelihood function of the CMBR
signal $C_B$ inside the bins is well approximated by an offset lognormal
distribution, such that the quantity $D_B=ln(C_B+x_B)$ (where $x_B$ is the
offset correction) is a gaussian variable\cite{BJK2K}. The likelihood for a
given cosmological model is then defined by $-2{\rm ln} L
=(D_B^{th}-D_B^{ex})M_{BB'}(D_{B'}^{th}-D_{B'}^{ex})$, where $M_{BB'}$ is a
matrix that defines the noise correlations and $D_B^{th}$ is the offset
lognormal theoretical band power. The theoretical $D_B^{th}$ were generated
using fast and accurate Boltzmann solvers \cite{selj}. Our database of
models is sampled in {\it physical variables} as in \cite{lange}, but we
only consider flat models, $h=0.65 {\pm} 0.2$ and the effective number of
neutrinos is allowed to vary up to $N_\nu = 20$. Following \cite{D&K} we
can therefore constrain the parameter of interest, $\Omega_Bh^2$ and
$N_{\nu}$, by finding the remaining ``nuisance'' parameters which maximize
them. As mentioned, at 95 and 99$\%$ C.L. we found $N_\nu \leq 13$ and
$N_\nu \leq 16$, respectively. A similar bound, using BOOMERanG data only,
has been obtained in \cite{Hannestad}.

In figure 2 we have summarized the second main result of our analysis. In
the $\Omega_B h^2-N_\nu$ plane we show the 95$\%$ C.L. likelihood regions
for both the high and low $D$ measurements, as well as the analogous
contours for standard BBN, obtained running our code with $\xi_e=0$. In the
same plot we show the 68 and 95 $\%$ C.L. regions obtained by CMBR data.

As a first comment, the standard BBN, $\xi_e=0$, and CMBR data analysis
lead to quite different values for $\Omega_B h^2$. This can be clearly seen
from the reported 95$\%$ results, but we have verified that the 99$\%$ C.L.
contour for high $D$ has no overlap with the region picked up by BOOMERanG
and MAXIMA-1 data, and a very marginal one for low $D$.
\begin{figure}[ht]
\begin{center}
\epsfxsize=6.3cm
\epsfysize=5cm
\epsffile{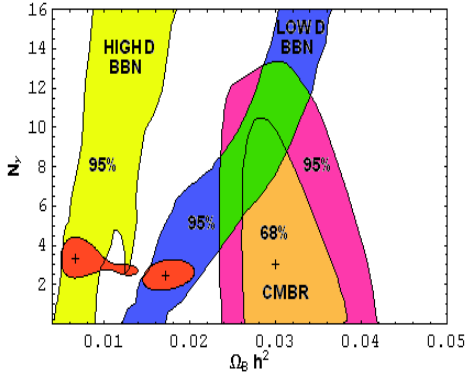}
\end{center}
\caption{The 95$\%$ C.L. contours in the $\Omega_B h^2-N_\nu$ plane
compatible with degenerate BBN (large bands) and standard BBN (small
regions) are plotted for both high $D$ (left) and low $D$ (right). The same
contours for 68$\%$ and 95$\%$ C.L. from BOOMERanG and MAXIMA-1 CMBR data
are also reported.}
\end{figure}
For the degenerate scenario, increasing $N_\nu$, the allowed intervals for
$\Omega_B h^2$ shift towards larger values. However the high $D$ values
require a baryon content of the universe energy density which is still too
low, $\Omega_B h^2 \leq 0.018$, to be in agreement with CMBR results. A
large overlap is instead obtained for the low $D$ case, whose preferred
$\Omega_B h^2$ span the range $0.012 \leq \Omega_B h^2 \leq 0.036$. As
expected, a larger $N_\nu$ helps in improving the agreement with the high
CMBR $\Omega_Bh^2$ value, but is important to stress that a large value for
$N_\nu$ is not preferred by the CMBR data alone, being, in this case, the
best fit $N_\nu \sim 3$. If we only consider the $95 \%$ overlap region we
get the following conservative bounds:
\be
4 \leq N_\nu \leq 13~~~,~~~~~~~ 0.024 \leq \Omega_B h^2 \leq 0.034 \pp
\ee
In this region $\xi_e$ varies in the range $0.07 \leq \xi_e \leq 0.43$. As
we said, values $N_\nu \geq 3$, as suggested from our analysis, can be
either due to weak interacting neutrino degeneracy, or rather to other
unknown relativistic degrees of freedom.

In conclusion, we have shown how a precision analysis of BBN and CMBR data
is able to tell us about possible new features of the cosmological model
describing the evolution of the universe. We have quantitatively discussed
how larger values for the baryonic matter content $\Omega_B h^2$ may be
reconciled with BBN predictions in a degenerate scenario. In this respect
the observed low value of deuterium more easily fits with the constraints
given by BOOMERanG and MAXIMA-1 data. The crucial aspect of our analysis is
that this agreement is realized with an effective neutrino degrees of
freedom larger than 4 at 95$\%$. This represents an indication in favour of
a degenerate neutrino background and/or new particle species contributing
to relativistic matter in the universe. As final remark, we should note
that our CMB analysis was restricted on a specific class of models with a
limited numbers of parameters. Including curvature, a gravity waves
background, or removing our priors on $h$ would not move the peak of our
likelihood on $\Omega_B h^2$, but would enlarge our C.L. more on the high
value side \cite{lange}. Including drastical deviations from the standard
model like topological defects \cite{bouchet}, decoherence in the
primordial fluctuations \cite{lewin} or assuming a less general
inflationary model \cite{sugiyama} would drastically change the conclusions
of our work. Each of these effect leaves a characteristic imprint on CMB,
so hopefully with new data available in the near future,  as well as
further analysis of the BOOMERanG and MAXIMA results, it will be possible
to severely scrutinize this result.

One of the author, A. Melchiorri would like to thank P. de Bernardis, P.
Ferreira, J. Silk and N. Vittorio for valuable comments and discussions;
G. Miele would like to thank CERN TH-Division for support.

\end{document}